\documentclass[%
 reprint,
nobibnotes,
bibnotes,
 amsmath,amssymb,
 aps,
]{revtex4-1}

\usepackage{multirow}
\usepackage{amsmath}
\usepackage{booktabs}
\usepackage{graphicx}% Include figure files
\graphicspath{{C:\xtemp}}
\usepackage{dcolumn}% Align table columns on decimal point
\usepackage{bm}% bold math
\begin{document}

% \preprint{APS/123-QED}

\title{Cellular Ability to Sense Spatial Gradients in the Presence of Multiple Competitive Ligands}% Force line breaks with \\

\author{Shu-Hao Liou}
 \email{hellont92@gmail.com}
\author{Chia-Chu Chen}
 \email{chiachu@phys.ncku.edu.tw}
 \affiliation{Department of Physics, National Cheng Kung University, Tainan, Taiwan 70101}
\date{\today}% It is always \today, today,

\begin{abstract}
Many eukaryotic and prokaryotic cells can exhibit remarkable sensing ability under small gradient of chemical compound. In this study, we approach this phenomenon by considering the contribution of multiple ligands to the chemical kinetics within Michaelis-Menten model. This work was inspired by the recent theoretical findings from Bo Hu \textit{et al.} [Phys. Rev. Lett. 105, 048104 (2010)], our treatment with practical binding energies and chemical potential provides the results which are consistent with experimental observations.
\begin{description}
\item[PACS numbers]
87.17.Jj, 05.40.-a, 87.10.Mn, 87.18.Tt
\end{description}
\end{abstract}

\pacs{87.17.Jj, 05.40.-a, 87.10.Mn, 87.18.Tt}% PACS, the Physics and Astronomy
\maketitle
%\tableofcontents
\section{\label{sec:level0}INTRODUCTION}
Cellular sensing ability is a general but critical biological function, it plays important roles in cancer sensing, wound healing,
embryogenesis and neuronal development \cite{begin}. This remarkable ability allows cell to gain necessary energy and nutrition and obtain information from other cells \cite{information}. In general, the size of the cell is only a few micrometers \cite{receptors} but it can discriminate the correct direction of the gradient when tiny variance of chemical concentration. The minimum fluctuation about detecting the concentration can be seen as the accuracy of the sensing ability, which was addressed by Berg and Purcell \cite{1limit} and then modified by R. Endres \textit{et al.} who considered the unoccupied time intervals of the ligands and receptors \cite{2limit}. Moreover, other researchers reported the physical limits of spatial sensing ability \cite{bo}, and then demonstrated the impossibility to increase the ellipse cell's sensing ability by enlarging the cell's body \cite{elli}.

According to the report from Bo Hu \textit{et al.} \cite{bo}, the accuracy to sense the ligand gradient direction will be increased dramatically by larger cell. Their results are based on the basic assumption that the thermodynamics equilibrium and the chemical kinetics are equivalent. The thermodynamic results are obtained in the frame work of canonical ensemble. The gradient sensing ability is calculated from the partition function which is completely determined by the energy of the Hamiltonian. The results of \cite{bo} are very interesting and intriguing in the sense that the sensing ability can be established in such a simple model. In this work we modify the model of \cite{bo} such that the correlation between energy and concentration can be relaxed. This is done by considering the dynamics of ligands. The chemical equilibrium of ligand and receptor provides a way to break the energy-concentration constraint by introducing chemical potentials for addressing the problem of concentrations. In this work we have re-analyzed the problem of sensing ability with the grand canonical ensembles with the contribution of ligands included. Furthermore it is also interesting to address this sensing problem in a more general environment in different ligands. In our model, the cell can distinguish distinct ligands with different chemical dissociation constants. This result can only be achieved by considering the binding energies of both ligands. Fortunately, our approach can also be extended to the analysis of this general environment.

In this report, the assumptions and results of \cite{bo} are discussed in section \ref{sec:level1}. Our approach will be
presented in section \ref{sec:level2}, and we will have a brief discussion on the experiment results related to our model. Section \ref{sec:level2-1} provides the analysis of multiple ligands system within Michaelis-Menten model. The conclusion is given in section \ref{sec:level3}.

\section{\label{sec:level1}APPLYING ISING MODEL IN SENSING PROBLEM}
To deal with the cellular sensing problem, Bo Hu \textit{et al.} treated the receptors as an Ising spin chain. From this model,
they could derive the asymptotic variances as functions of the gradient steepness $p$ and direction of gradient $\phi$ which are
parameters related to the concentration. According to the Cram\'{e}r-Rao inequality, these variances determine the lowest uncertainties of cell sensing ability \cite{distri}. Here we briefly review their calculation procedures.
% 4: ref 18; 5:the comment

In this model, the cell with diameter $L$ and $N$ receptors were immersed in the concentration environment which contains identical ligands. In their work all results were calculated with $N=80000$, which is close to the practical situation \cite{receptor}. The local concentration of ligands at $n$th receptor is $C_n=C_0\exp[\frac{p}{2}\cos(\varphi_n-\phi)]$, where $C_0$ is the background concentration, $p$ is the steepness of gradient $(p\equiv\frac{L}{C_0}|\vec{\bigtriangledown}C|)$, $\varphi_n=2n\pi/N$ denoted the location of $n$th receptor and direction of gradient is $\phi$. In this approach, ignoring the dynamics of ligands, the system is completely described by receptors which have only binding state ($S_n=+1$) with energy $-\varepsilon_n$ and unbinding state ($S_n=-1$) with energy $\varepsilon_n$, where $\varepsilon_n$ is given in the unit of thermal energy $k_BT$. Due to the Boltzmann distribution, the binding probability of the $n$th receptor is $P_{on}=\frac{e^{\varepsilon_n}}{e^{\varepsilon_n}+e^{-\varepsilon_n}}$.
Using simple receptor-ligand kinetics, the binding probability of the $n$th receptor is $P_{on}=\frac{C_n}{C_n+K_d}$ where
$K_d$ is the dissociation constant. By assuming these probabilities are identical, the free energy is
\begin{equation}
\varepsilon_n=\frac{1}{2}\ln\frac{C_0}{K_d}+\frac{p}{4}\cos(\varphi_n-\phi).
\label{eq:one}
\end{equation}
By defining three statistical quantities $(z_0, z_1, z_2)=(\sum_nS_n, \sum_n{\frac{1}{2}S_n\cos{\varphi_n}}, \sum_n{\frac{1}{2}S_n\sin{\varphi_n}})$ and transformation $(\alpha_0, \alpha_1, \alpha_2)=(\frac{1}{2}\ln\frac{C_0}{K_d}, p\cos\phi, p\sin\phi)$, the Hamiltonian is given as $H_N\{S_n\}=-\alpha_0z_0-\frac{z_1\alpha_1+z_2\alpha_2}{2}$ and by computing the logarithm partition function $\ln{Q_N}=N\ln(2\cosh{\alpha_0})+\frac{Np^2}{64\cosh^2{\alpha_0}}+\mathcal{O}(p^4)$, they obtained the expectation values and fluctuations of $z_1$ and $z_2$:
\begin{equation}
<z_{1,2}>=\frac{NC_0K_d\alpha_{1,2}}{4(C_0+K_d)^2}+\mathcal{O}(p^3),
\label{eq:two}
\end{equation}
\begin{equation}
Var[z_{1,2}]=\frac{NC_0K_d}{2(C_0+K_d)^2}+\mathcal{O}(p^2).
\label{eq:three}
\end{equation}
Under the assumption $Cov[z_1,z_2]=0$ \cite{bo} the joint probability density of $z_1$ and $z_2$ is the Gaussian function, the
asymptotic variances $\sigma^2_p$ and $\sigma^2_{\phi}$ can be obtained as:
\begin{equation}
\sigma^2_p=\frac{8(C_0+K_d)^2}{NC_0K_d},
\label{eq:four}
\end{equation}
\begin{equation}
\sigma^2_\phi=\frac{8(C_0+K_d)^2}{NC_0K_dp^2}.
\label{eq:five}
\end{equation}
These asymptotic variances are the minimum fluctuations with unbiased estimation of $p$ and $\phi$ which are related to sensing
ability. Therefore, the fluctuation $\sigma^2_{\phi}$ drops as gradient increase, and the sensing ability will be increased with enlarged cell volume since $p\propto{L}$ and $N\propto{L^\delta}$ with $0\leq\delta\leq2$ \cite{bo}.

\begin{figure}
\includegraphics[width=0.35\textwidth]{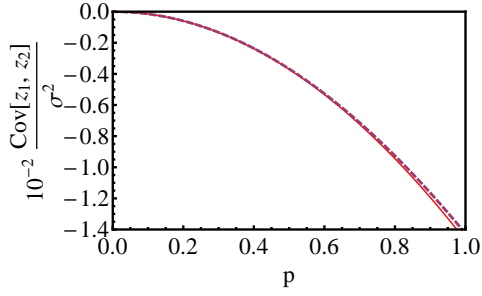}% Here is how to import EPS art
\caption{\label{fig:one} (Color online). The $Cov[z_1,z_2]$ of both models under $K_d=C_0$ and $\phi=\pi/5$ arbitrarily. The dashed line represents $Cov[z_1,z_2]/\sigma^2$ versus $p$ in \cite{bo}, and solid line represents $\frac{Cov[z_1,z_2]}{\sqrt{<\Delta(z_1)^2><\Delta(z_2)^2>}}$ with $p$ in our model which will be described in next section.}
\end{figure}

In passing we would like to show the detailed analysis about the covariance $Cov[z_1,z_2]=0$ for justifying the joint probability density as Gaussian distribution. The $Cov[z_1,z_2]$ is given by:
\begin{equation}
Cov[z_1,z_2]=-\sum^N_n\frac{1}{4}\cos{\varphi_n}\sin{\varphi_n}\frac{(e^{\varepsilon_n}-e^{-\varepsilon_n})^2}{(e^{\varepsilon_n}+e^{-\varepsilon_n})^2},
\label{eq:seven}
\end{equation}
and the numerical results is plotted in FIG. \ref{fig:one}, in which one can see that it is very small and can be approximated to zero, such that the probability density can be described by Gaussian distribution. Even if our model will use this approximation for further analyzing and $Cov[z_1,z_2]$ in our model is also shown in FIG. \ref{fig:one}.

\section{\label{sec:level2}ADDING THE LIGANDS INFORMATION}

To calculate the physical limit of multi-ligands system, the ligand's concentration should be considered in the partition function. The grand canonical ensemble is then appropriate to construct our model. Here we adapted the notations of \cite{bo} where the cell with size $L$ has $N$ receptors, and the spatial information of ligands are given by the concentration $C_n$, gradient steepness $p$
and the direction of the gradient $\phi$. Each receptor should only sense the ligands inside an identical sensing
volume independently. In other words, the $n$th receptor's sensing
volume $v$ is completely separate with others'. Let
$L_n\equiv{vC_n}$ denotes the number of ligand which can be sensed
only by the $n$th receptor. The approximative expression of sensing
volume is $v=4\pi{r^2}d/N$, where $r=L/2$ is the radius of cell and
$d$ is the size of ligand. However, to be consistent in this model, three different energy levels were set up to describe different
states: the unbinding energy $\varepsilon_u$, binding energy
$\varepsilon_b$ and ligand energy $\varepsilon_l$; and the
corresponding chemical potentials are denoted by $\mu_u$, $\mu_b$
and $\mu_l$ respectively. We notice that instead of being position dependent as resulted in \cite{bo}, these energy levels are position independent as they should be according to basic quantum principle. The Hamiltonian of this system is
\begin{eqnarray}
H_N\{S_n\}=&&\sum_{n=1}^N[\varepsilon_b(\frac{1}{2}+\frac{S_n}{2})+\varepsilon_u(\frac{1}{2}-\frac{S_n}{2}) \nonumber\\
&&+\varepsilon_l(L_n-\frac{1}{2}-\frac{S_n}{2})].
\label{eq:eight}
\end{eqnarray}

\begin{figure}
\includegraphics[width=0.51\textwidth]{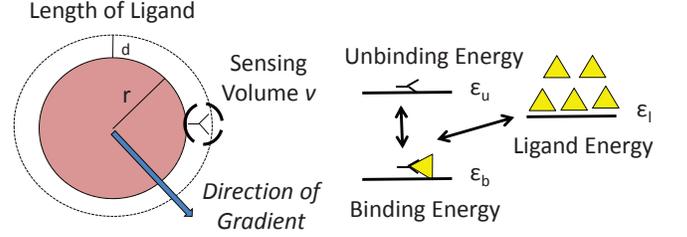}% Here is how to import EPS art
\caption{\label{fig:oneonw} (Color online). The diagrams of our model: The cell with identical receptors is located in concentration pool with particular direction of gradient. Each receptor can only sense the ligands inside the sensing volume $v$. Three different energy levels with binding, unbinding and ligand states are all independent of location. }
\end{figure}

To simplify our discussion, we denote the $n$th receptor's grand
canonical partition function as the binding part $z_{bn}$ and unbinding
part $z_{un}$, which are given by:
\begin{subequations} \label{eq:wholezero}
\begin{equation}
z_{bn}=\frac{1}{(L_n-1)!}e^{-\beta[(\varepsilon_b-\mu_b)+(\varepsilon_l-\mu_l)(L_n-1)]}, \label{subeq:01}
\end{equation}
\begin{equation}
z_{un}=\frac{1}{L_n!}e^{-\beta[(\varepsilon_u-\mu_u)+(\varepsilon_l-\mu_l)L_n]}. \label{subeq:02}
\end{equation}
\end{subequations}
The factorial factors appear here since ligands are all identical.
The total grand canonical partition function of the whole cell
becomes
\begin{widetext}
\begin{equation}
\mathcal{Z}=\prod^N_{n=1}(z_{bn}+z_{un})=\prod^N_{n}\sum_{S_n=\{+1,-1\}}\frac{1}{(L_n-\frac{1}{2}-\frac{S_n}{2})!}e^{-\beta[(\varepsilon_b-\mu_b)(\frac{1}{2}+\frac{S_n}{2})+(\varepsilon_u-\mu_u)(\frac{1}{2}-\frac{S_n}{2})+(\varepsilon_l-\mu_l)(L_n-\frac{1}{2}-\frac{S_n}{2})]}.
\label{eq:eleven}
\end{equation}
\end{widetext}
The binding probability due to the Boltzmann distribution is
$P_{bn}=\frac{z_{bn}}{z_{bn}+z_{un}}$. The binding
probability of chemical equilibrium at the $n$th receptor is given
by $P_{cn}=\frac{C_n}{C_n+K_d}$. By imposing $P_{bn}=P_{cn}$ we obtain the relation:
\begin{equation}
a\equiv\frac{K_d}{C_0}=\frac{1}{vC_0}e^{-\beta[-(\varepsilon_b-\mu_b)+(\varepsilon_u-\mu_u)+(\varepsilon_l-\mu_l)]}.
\label{eq:12}
\end{equation}
Eq. (\ref{eq:12}) shows that the chemical dissociation constant $K_d$ under this
assumption should depend on energy levels, chemical potentials and
the sensing volume. Moreover, the binding probability would depend
on concentration and position, but the energy levels are independent of position. By using Eq.
(\ref{eq:12}) all the dependence on $\varepsilon$ and $\mu$ are
replaced by dissociation constant $K_d$ and local concentration $C_0$. In particular, the sensing ability will
be determined by $a\equiv{K_d/C_0}$ and more details will be discussed as the following.

With the statistical quantities
$(z_0,z_1,z_2)=(\sum_nS_n,\sum_n{\frac{1}{2}S_n\cos{\varphi_n}},\sum_n{\frac{1}{2}S_n\sin{\varphi_n}})$ and the transformation $(\alpha_1,\alpha_2)=(p\cos\phi,p\sin\phi)$,
the analysis of discrimination can be proceeded with the expectation
values $\overline{z_1}$ and $\overline{z_2}$. These expectation
values are calculated by using direct summation method. Defining $\zeta_n=e^{i\varphi_n}$, then
$\overline{z_1}=Re[\sum_n^N\zeta_n\frac{(z_{bn}-z_{un})}{(z_{bn}+z_{un})}]$
and
$\overline{z_2}=Im[\sum_n^N\zeta_n\frac{(z_{bn}-z_{un})}{(z_{bn}+z_{un})}]$.
Under the small $p$ assumption (which is also true for experimental
environment), we expand the summation to second order of $p$, and
treat the summation as an integral over $[0,2\pi]$, hence the integrals can be computed as
\begin{widetext}
\begin{subequations}\label{eq2:1}
\begin{equation}
\overline{z_1}\simeq\frac{N}{2\pi}\int^{2\pi}_0\frac{\cos\varphi}{1+a[1-(\frac{\alpha_1}{2}\cos\varphi+\frac{\alpha_2}{2}\sin\varphi)+\frac{1}{2}(\frac{\alpha_1}{2}\cos\varphi+\frac{\alpha_2}{2}\sin\varphi)^2]}d\varphi,
\label{eq2:1:1}
\end{equation}
\begin{equation}
\overline{z_2}\simeq\frac{N}{2\pi}\int^{2\pi}_0\frac{\sin\varphi}{1+a[1-(\frac{\alpha_1}{2}\cos\varphi+\frac{\alpha_2}{2}\sin\varphi)+\frac{1}{2}(\frac{\alpha_1}{2}\cos\varphi+\frac{\alpha_2}{2}\sin\varphi)^2]}d\varphi,
\label{eq2:1:2}
\end{equation}
\end{subequations}
\end{widetext}
and both results of $\overline{z_1}=Re[\Omega]$ and $\overline{z_2}=Im[\Omega]$ are:
\begin{subequations} \label{eq:wholeone}
\begin{equation}
\Omega=\frac{8Ne^{i\phi}}{ap^2\sqrt{A\frac{(2+a)}{a}}}(\sqrt{\frac{2+a}{a}}\cos{\frac{\theta}{2}}+\sin{\frac{\theta}{2}}), \label{subeq:11}
\end{equation}
\begin{equation}
Ae^{i\theta}=(-1-\frac{8}{ap^2})+i\frac{8}{p^2}\sqrt{\frac{(2+a)}{a}}. \label{subeq:12}
\end{equation}
\end{subequations}
The fluctuations of $z_1$ and $z_2$ are
$\sigma^2_1\equiv<\Delta(z_1)^2>=Re[\sigma^2_{+}]$ and
$\sigma^2_2\equiv<\Delta(z_2)^2>=Re[\sigma^2_{-}]$ where
\begin{widetext}
\begin{eqnarray}
\sigma^2_\pm=&&\frac{2N}{p\sqrt{a(2+a)}}\bigg\{\frac{\sin\frac{\theta}{2}}{2\sqrt{A}}\pm{e^{i2\phi}}\big(\frac{4}{p}\sqrt{\frac{2+a}{a}}-\sqrt{A}\sin\frac{\theta}{2}-\frac{1}{\sqrt{A}}(\frac{8}{p^2}\sqrt{\frac{2+a}{a}}\cos\frac{\theta}{2}+\frac{8}{ap^2}\sin\frac{\theta}{2})\big)\bigg\}+ \nonumber \\
&&\frac{N}{p\sqrt{a(1+a)^3}}\bigg\{2\frac{\pm{e^{2i\phi}}-1}{\eta_2}\sin\lambda_2\pm{e^{2i\phi}}\big(\frac{\eta^2_1}{\eta_2}\sin(2\lambda_1-\lambda_2)-\eta_1\sin\lambda_1\big)\bigg\},
\end{eqnarray}
\end{widetext}
with
\begin{subequations} \label{eq:wholetwo}
\begin{equation}
\eta_1e^{i\lambda_1}=-\frac{4(a+1)}{p(1+2a)}(1+i\sqrt{\frac{1+a}{a}}), \label{subeq:21}
\end{equation}
\begin{equation}
\eta_2e^{i\lambda_2}=\sqrt{[\frac{-1}{ap^2}(\frac{4a+4}{(1+2a)})^2-4]+2i[(\frac{4a+4}{p(1+2a)})^2\sqrt{\frac{1+a}{a}}]}. \label{subeq:22}
\end{equation}
\end{subequations}

We should mention that all the results are real, although complex numbers appear from complex integral calculations to simplify the
notation. It is interesting to note that even though the grand
canonical ensemble depends on energy level $\varepsilon$ and
chemical potential $\mu$, in this model the expectation values and
fluctuations only depend on $p$, $\phi$ and $a$, where $a$
implicitly depends on $\varepsilon$ and $\mu$ as given in Eq.
(\ref{eq:12}). The detail dynamics of the system such as the energy
levels and the chemical potentials determine the dissociation
constant $K_d$. However, the sensing ability relies on the detail
dynamics only through the specification of $a$. In \cite{bo} they used
the procedures given in \cite{distri} to estimate the joint
probability as Gaussian function. To apply this assumption one
should check the value of $Cov[z_1,z_2]$ explicitly. By using
Eq.(\ref{eq:seven}) the correlation which was divided by fluctuation
is obtained in FIG. \ref{fig:one}, where one can see clearly that $Cov[z_1,z_2]\simeq{0}$ as $p$ being small. Under this assumption, according to
\cite{distri} one can assume the joint probability density as
Gaussian function, i.e.,
$f(z_{1,2}|\overline{z_{1,2}})\simeq\frac{1}{2\pi\sigma_1\sigma_2}\exp[-\frac{(z_1-\overline{z_1})^2}{2\sigma_1^2}-\frac{(z_2-\overline{z_2})^2}{2\sigma_2^2}]$.
However, for large $p$, it seems that a more detailed analysis is called for. Under
the Maximum Likelihood Estimator (MLE) theorem, the estimators
of $p$ and $\phi$, denoted by $\sigma_p^2$ and $\sigma_\phi^2$ respectively, can be obtained near the expectation value without
any bias \cite{distri}. The estimators ( also asymptotic variances) related to the sensing ability can be derived by the
Cram\'{e}r-Rao lower bound:
\begin{subequations} \label{eq:wholeaddone}
\begin{eqnarray}
&&\sigma_p^2=\frac{1}{<(\partial_p\ln{f})^2>}=\label{subeq:add1}\\
&&\frac{1}{\frac{1}{2\sigma_1^4}(\frac{\partial\sigma_1^2}{\partial{p}})^2+\frac{1}{2\sigma_2^4}(\frac{\partial\sigma_2^2}{\partial{p}})^2+\frac{1}{\sigma_1^2}(\frac{\partial{\overline{z_1}}}{\partial{p}})^2+\frac{1}{\sigma_2^2}(\frac{\partial{\overline{z_2}}}{\partial{p}})^2}, \nonumber
\end{eqnarray}
\begin{eqnarray}
&&\sigma_\phi^2=\frac{1}{<(\partial_{\phi}\ln{f})^2>}=\label{subeq:add2}\\
&&\frac{1}{\frac{1}{2\sigma_1^4}(\frac{\partial\sigma_1^2}{\partial{\phi}})^2+\frac{1}{2\sigma_2^4}(\frac{\partial\sigma_2^2}{\partial{\phi}})^2+\frac{1}{\sigma_1^2}(\frac{\partial{\overline{z_1}}}{\partial{\phi}})^2+\frac{1}{\sigma_2^2}(\frac{\partial{\overline{z_2}}}{\partial{\phi}})^2}. \nonumber
\end{eqnarray}
\end{subequations}

\begin{figure}
\includegraphics[width=0.4\textwidth]{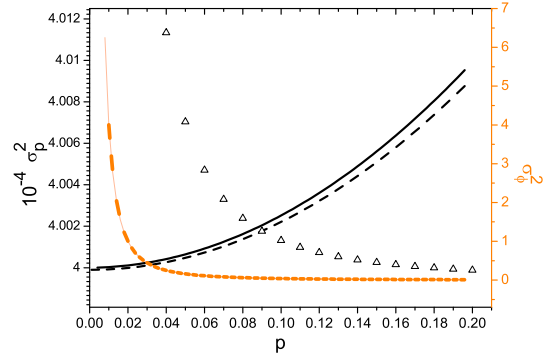}\\% Here is how to import EPS art
\caption{\label{fig:four} (Color online). The direct summation results of $\sigma_p^2$ (black) and $\sigma_\phi^2$ (orange) versus $p$ with $a=1.0$ or $K_D=C_0$. The dash line represents the results in \cite{bo}. The triangle represents the $\sigma_\phi^2$ under $a=0.01$, which means the local concentration is much larger than the dissociation constant $K_d$.}
\end{figure}

Following figures are the computational results without any approximation. FIG. \ref{fig:four} shows that $\sigma^2_{\phi}$ which relates to the ability of sensing will decrease by increasing $p$. Since the steepness of gradient $p$ is proportional to the size of cell $L$, the sensing ability would increase dramatically when volume of the cell expands. The result shows similar conclusion as \cite{bo} without $\varepsilon_n$ being $C_n$ and position dependent. In FIG. \ref{fig:four}, one can see that $\sigma_p^2$ almost remains the same for different $p$, however $\sigma_\phi^2$ will decrease when the steepness increases under the condition $K_d=C_0$. We also plotted their results \cite{bo} in FIG. \ref{fig:four}, one can see that the result of two models are quiet close with similar characteristics.

It is interesting to see our results are related to experiment observations. It is well known that the sensing ability or sensing accuracy strongly depends on the steepness of gradient but weakly correlates to background concentration \cite{steepness}.
In this work we have analyzed this aspect and the results are also plotted in FIG. \ref{fig:four}. $\sigma^2_{\phi}$ approaches to
zero when $p$ is large, which indicates good sensing ability in this range. When $p$ is larger than $0.1$, it can be seen that $\sigma^2_{\phi}$ of different concentrations ($a=0.01$ and $a=1000$) are more or less the same, which means that the local
concentration will weakly depend on sensing ability when $p>0.2$. However, when $p$ is less than $0.1$, the effect of concentration might be large for sensing ability.

\begin{figure}
\includegraphics[width=0.4\textwidth]{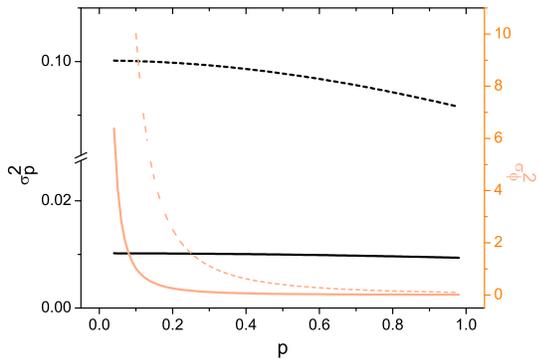}\\% Here is how to import EPS art
\caption{\label{fig:six} (Color online). The results of $\sigma_p^2$ (black) and $\sigma_\phi^2$ (light orange) versus $p$ in different background environment. The dash line represents $a=0.01$ and the solid line shows the result of $a=1000$. The result is plotted by direct summation without any approximation.}
\end{figure}

Moreover, it is known that cells can show remarkable sensing ability in particular steepness and concentration range. For instance, \textit{Dictyostelium} cells will move toward cyclic adenosine 3',5'-monophosphate (cAMP) to function as a
chemoattractant \cite{camp}. In general, under $p=2\%$ and $K_d\sim100nM$, the cell can exhibit the sensing ability when cAMP
is in a range between $10 pM$ to $10 \mu{M}$ \cite{range1}, or $a=0.01\sim1000$ in our model. FIG. \ref{fig:six} shows
the two asymptotic variances  $\sigma_p^2$ and  $\sigma_\phi^2$ within this range. It is obvious that the fluctuation is quite large
at $a=1000$ (which means small background concentration or large disassociate constant), but the $\sigma_\phi^2$ is below $1.0$ when
$p>0.32$. In other words, the sensing ability is exhibited under small concentration and small steepness of gradient.

\section{\label{sec:level2-1}MULTIPLE LIGANDS WITH COMPETITIVELY BINDING IN CELLULAR SENSING ABILITY}
In practical biological systems, many receptors can bind with different ligands on the same site to conduct many important functions \cite{multiple}. For example, the ions $\mbox{H}^{+}$, $\mbox{K}^+$ and $\mbox{Mg}^{2+}$ can bind with eukaryotic cell's $\mbox{Ca}^{2+}$-binding sites of calmodulin \cite{ca}, which are related to the intracellular movement, metabolism and apoptosis \cite{cafunction}. Moreover, different ligands might present dissimilar effect after binding. For instant, platelet-derived growth factor (PDGF), which is excreted by platelet $\alpha$-granules during the injury, can strongly attract monocytes and neutrophils\cite{competitive}. However, the protamine sulfate which competitively binds to the surface of monocyte and neutrophil will block the chemotaxis \cite{competitive} and shows distinct role in chemotaxis. Therefore it is interesting to analyze the sensing ability in the multiple ligands environment.

In this section we consider the concentration environment with two different ligands denoted by Ligand 1 and Ligand 2, which can bind on the same site of the receptor whenever it is unoccupied by any ligand. The receptors are still be treated as Ising spin chain. Since the effects of binding for these ligands are in general distinct, two different operators $S_{1n}$ and $S_{2n}$ are needed to describe the states of receptor for Ligand 1 and Ligand 2 respectively. The eigenvalues of both operators are equivalent, such as binding state ($+1$) and unbinding state ($-1$). The parameters of two ligands are listed in TABLE \ref{tab:table1}, where the concentrations of both ligands are $C_n=C_0\exp[\frac{p_1}{2}\cos(\varphi_n-\phi_1)]$ and $D_n=D_0\exp[\frac{p_2}{2}\cos(\varphi_n-\phi_2)]$ at the $n$th receptor, and the energy level and chemical potential for unbinding receptor are $\varepsilon_{u}$ and $\mu_{u}$ respectively. All the energy levels and chemical potentials of the system are independent of position and concentration. The Hamiltonian of the system is expressed as:
\begin{widetext}
\begin{eqnarray}
H_N\{S_{1n},S_{2n}\}=&&\sum_{n=1}^N[-(\frac{(S_{1n}+1)(S_{2n}-1)}{4})\varepsilon_1'-(\frac{(S_{2n}+1)(S_{1n}-1)}{4})\varepsilon_2'-(\frac{(S_{1n}-1)(S_{2n}-1)}{4}) \varepsilon_u \nonumber\\
&&+(vC_n+\frac{(S_{1n}+1)(S_{2n}-1)}{4})\varepsilon_1+(vD_n+\frac{(S_{2n}+1)(S_{1n}-1)}{4})\varepsilon_2].
\label{eq:hetero0}
\end{eqnarray}
\end{widetext}

\begin{table}
\caption{\label{tab:table1} The parameters for two ligands at the $n$th receptor. }
\begin{ruledtabular}
\begin{tabular}{lcc}
 Type & Ligand 1& Ligand 2\\
\colrule
 %Concentration & $C_n$ & $D_n$\\
 Background Concentration &$C_0$ & $D_0$\\
 Steepness of Gradient & $p_1$ & $p_2$\\
 Direction of Gradient & $\phi_1$ & $\phi_2$\\
 Energy Level of Unbinding Ligand & $\varepsilon_1$ & $\varepsilon_2$\\
 Energy Level of Binding Ligand& $\varepsilon_1'$ & $\varepsilon_2'$\\
 Chemical Potential of Unbinding Ligand& $\mu_1$ & $\mu_2$\\
 Chemical Potential of Binding Ligand& $\mu_1'$ & $\mu_2'$\\
\end{tabular}
\end{ruledtabular}
\end{table}

The grand canonical partition function at the $n$th site $Z_n$  can be separated as parts of unbinding ($z_{un}$), binding with Ligand 1 ($z_{1n}$) and Ligand 2 ($z_{2n}$):
\begin{subequations} \label{eq:hetero1}
\begin{equation}
z_{\scriptstyle{{un}}}=\frac{1}{(vC_n)!(vD_n)!}e^{-\beta[(\varepsilon_u-\mu_u)+(\varepsilon_1-\mu_1)vC_n+(\varepsilon_2-\mu_2)vD_n]}, \label{subeq:h11}
\end{equation}
\begin{eqnarray}
&&z_{1n}=\frac{1}{(vC_n-1)!(vD_n)!}\times \nonumber\\
&&e^{-\beta[(\varepsilon_1'-\mu_1')+(\varepsilon_1-\mu_1)(vC_n-1)+(\varepsilon_2-\mu_2)vD_n]}, \label{subeq:h12}
\end{eqnarray}
\begin{eqnarray}
&&z_{2n}=\frac{1}{(vC_n)!(vD_n-1)!}\times \nonumber\\
&&e^{-\beta[(\varepsilon_2'-\mu_2')+(\varepsilon_1-\mu_1)vC_n+(\varepsilon_2-\mu_2)(vD_n-1)]}. \label{subeq:h13}
\end{eqnarray}
\end{subequations}
The total partition function $\Xi$ for the cell therefore becomes
\begin{equation}
\Xi=\prod^N_{n=1}{Z_n}=\prod^N_{n=1}(z_{un}+z_{1n}+z_{2n}).
\label{eq:hetero2}
\end{equation}

\begin{table}[h]
\caption{\label{tab:table2} The parameters of Michaelis-Menten Model at the $n$th receptor. Abbreviation: IC, initial concentration; FC, final concentration after equilibrium; CDC, chemical dissociation constant.}
\begin{ruledtabular}
\begin{tabular}{lc}
 Type & Parameter \\
\colrule
 IC of Enzyme & $e_{n0}$ \\
 FC of Unbinding Enzyme &$e_n$\\
 FC of Enzyme Bind with Ligand 1 & $c_{n1}$\\
 FC of Enzyme Bind with Ligand 2& $c_{n2}$\\
 CDC for Ligand 1 and Enzyme & $K_1\equiv{k_{\bar{1}}}/k_1$\\
 CDC for Ligand 2 and Enzyme & $K_2\equiv{k_{\bar{2}}}/k_2$\\
\end{tabular}
\end{ruledtabular}
\end{table}

To identify the probabilities of binding and unbinding cases in chemical kinetics, we apply the Michaelis-Menten Model which is widely used in non-allosteric enzymes \cite{biochemistry} where the receptors are treated as isolated enzymes in this model. The parameters of Michaelis-Menten Model are listed in Table \ref{tab:table2}. The main equations for competitive constrain and chemical equilibrium for Ligand 1 and Ligand 2 can be written as:
\begin{subequations} \label{eq:hetero3}
\begin{equation}
e_{n0}=e_n+c_{n1}+c_{n2}, \label{subeq:h31}
\end{equation}
\begin{equation}
\frac{dc_{n1}}{dt}=k_1e_nC_n-k_{\bar{1}}c_{n1}=0, \label{subeq:h32}
\end{equation}
\begin{equation}
\frac{dc_{n2}}{dt}=k_2e_nD_n-k_{\bar{2}}c_{n2}=0. \label{subeq:h33}
\end{equation}
\end{subequations}

At equilibrium, one can solve for the final concentrations $c_{n1}$, $c_{n2}$ and the unbinding receptors. The ratios of these variables and the original concentration of receptor can be seen as the probability of binding or unbinding state at $n$th site. Therefore, the probabilities of binding with Ligand 1 ($P_{1n}$), Ligand 2 ($P_{2n}$) or without binding ($P_{un}$) are:
\begin{widetext}
\begin{subequations} \label{eq:hetero4}
\begin{equation}
P_{un}=\frac{e_n}{e_{n0}}=\frac{1}{1+\frac{C_n}{K_1}+\frac{D_n}{K_2}}\equiv\frac{z_{un}}{Z_n}=\frac{1}{1+vC_ne^{-\beta[(\varepsilon_1'-\mu_1')-(\varepsilon_1-\mu_1)-(\varepsilon_{u}-\mu_{u})]}+vD_ne^{-\beta[(\varepsilon_2'-\mu_2')-(\varepsilon_2-\mu_2)-(\varepsilon_{u}-\mu_{u})]}}, \label{subeq:h41}
\end{equation}
\begin{equation}
P_{1n}=\frac{c_{n1}}{e_{n0}}=\frac{\frac{C_n}{K_1}}{1+\frac{C_n}{K_1}+\frac{D_n}{K_2}}\equiv\frac{z_{1n}}{Z_n}=\frac{vC_ne^{-\beta[(\varepsilon_1'-\mu_1')-(\varepsilon_1-\mu_1)-(\varepsilon_{u}-\mu_{u})]}}{1+vC_ne^{-\beta[(\varepsilon_1'-\mu_1')-(\varepsilon_1-\mu_1)-(\varepsilon_{u}-\mu_{u})]}+vD_ne^{-\beta[(\varepsilon_2'-\mu_2')-(\varepsilon_2-\mu_2)-(\varepsilon_{u}-\mu_{u})]}},  \label{subeq:h42}
\end{equation}
\begin{equation}
P_{2n}=\frac{c_{n2}}{e_{n0}}=\frac{\frac{D_n}{K_2}}{1+\frac{C_n}{K_1}+\frac{D_n}{K_2}}\equiv\frac{z_{2n}}{Z_n}=\frac{vD_ne^{-\beta[(\varepsilon_2'-\mu_2')-(\varepsilon_2-\mu_2)-(\varepsilon_{u}-\mu_{u})]}}{1+vC_ne^{-\beta[(\varepsilon_1'-\mu_1')-(\varepsilon_1-\mu_1)-(\varepsilon_{u}-\mu_{u})]}+vD_ne^{-\beta[(\varepsilon_2'-\mu_2')-(\varepsilon_2-\mu_2)-(\varepsilon_{u}-\mu_{u})]}}. \label{subeq:h43}
\end{equation}
\end{subequations}
\end{widetext}
Similar to the case of single ligand system, we introduce $b_1$ and $b_2$:
\begin{subequations} \label{eq:hetero5}
\begin{equation}
b_1\equiv\frac{K_1}{C_0}=\frac{1}{vC_0}e^{-\beta[-(\varepsilon_1'-\mu_1')+(\varepsilon_1-\mu_1)+(\varepsilon_{u}-\mu_{u})]}, \label{subeq:h51}
\end{equation}
\begin{equation}
b_2\equiv\frac{K_2}{D_0}=\frac{1}{vD_0}e^{-\beta[-(\varepsilon_2'-\mu_2')+(\varepsilon_2-\mu_2)+(\varepsilon_{u}-\mu_{u})]}. \label{subeq:h52}
\end{equation}
\end{subequations}

It is necessary to define the statistical parameters for Ligands, which are $(x_1,x_2)=(\sum_n{\frac{1}{2}S_{1n}\cos{\varphi_n}},\sum_n{\frac{1}{2}S_{1n}\sin{\varphi_n}})$, and $(x_3,x_4)=(\sum_n{\frac{1}{2}S_{2n}\cos{\varphi_n}},\sum_n{\frac{1}{2}S_{2n}\sin{\varphi_n}})$ for Ligand 1 and 2 respectively. Furthermore, the expectation value for each parameter is $\overline{x}_{i}\equiv{<x_i>}$, and it's fluctuation can be calculated by $\sigma^2_i\equiv{<(x_i-\overline{x}_i)^2>}$. Assuming these parameters are independent, their probability distributions can be described by Gaussian function $g\simeq\Pi^4_{i=1}\frac{1}{2\pi\sigma_i}\exp[-\frac{(x_i-\overline{x}_i)^2}{\sigma^2_i}]$, then the Fisher information matrix $\textbf{F}$ then can be obtained as \cite{distri}:

\begin{widetext}
\begin{eqnarray}
\textbf{F}= (-1)\left(
\begin{array}{cccc}
<\frac{\partial^2\ln{g}}{\partial{\overline{x}_1}^2}> &<\frac{\partial^2\ln{g}}{\partial{\overline{x}_1}\partial{\overline{x}_2}}> &<\frac{\partial^2\ln{g}}{\partial{\overline{x}_1}\partial{\overline{x}_3}}> &<\frac{\partial^2\ln{g}}{\partial{\overline{x}_1}\partial{\overline{x}_4}}>\\
<\frac{\partial^2\ln{g}}{\partial{\overline{x}_1}\partial{\overline{x}_2}}> &<\frac{\partial^2\ln{g}}{\partial{\overline{x}_2}^2}>  &<\frac{\partial^2\ln{g}}{\partial{\overline{x}_2}\partial{\overline{x}_3}}> &<\frac{\partial^2\ln{g}}{\partial{\overline{x}_2}\partial{\overline{x}_4}}>\\
<\frac{\partial^2\ln{g}}{\partial{\overline{x}_1}\partial{\overline{x}_3}}> &<\frac{\partial^2\ln{g}}{\partial{\overline{x}_2}\partial{\overline{x}_3}}>  &<\frac{\partial^2\ln{g}}{\partial{\overline{x}_3}^2}>  &<\frac{\partial^2\ln{g}}{\partial{\overline{x}_3}\partial{\overline{x}_4}}> \\
<\frac{\partial^2\ln{g}}{\partial{\overline{x}_1}\partial{\overline{x}_4}}> &<\frac{\partial^2\ln{g}}{\partial{\overline{x}_2}\partial{\overline{x}_4}}> &<\frac{\partial^2\ln{g}}{\partial{\overline{x}_3}\partial{\overline{x}_4}}> &<\frac{\partial^2\ln{g}}{\partial{\overline{x}_4}^2}>
\end{array}\right).\label{eq:13}
\end{eqnarray}
\end{widetext}

\begin{figure}
\includegraphics[width=0.4\textwidth]{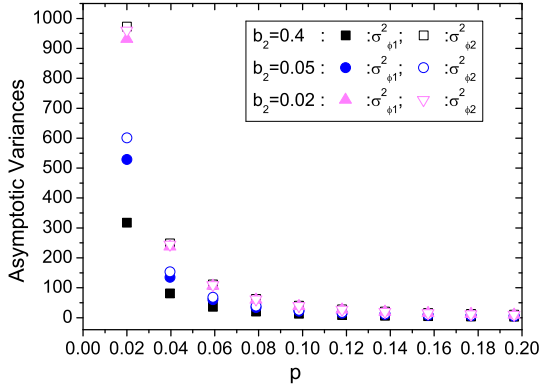}\\% Here is how to import EPS art
\caption{\label{fig:seven} (Color online). The results of $\sigma_{\phi_1}^2$ and $\sigma_{\phi_2}^2$ versus $p$ for $b_1=0.01$ with different $b_2$ with $\phi_1=0$ and $\phi_2=\pi/10$. We just plot the range $0.02\sim0.2$ due to the lack of space.}
\end{figure}

\begin{figure}
\includegraphics[width=0.4\textwidth]{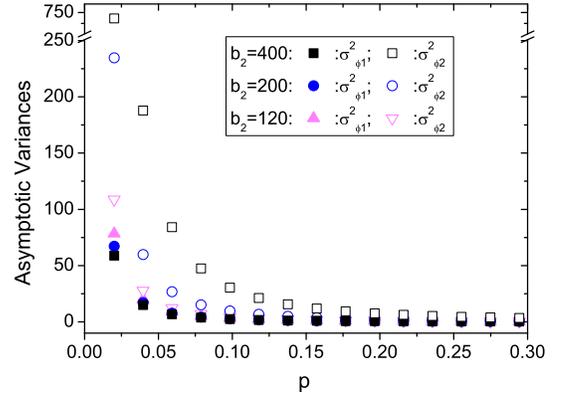}\\% Here is how to import EPS art
\caption{\label{fig:eight} (Color online). The results of $\sigma_{\phi_1}^2$ and $\sigma_{\phi_2}^2$ versus $p$ for $b_1=100$ and various $b_2$ with $\phi_1=0$ and $\phi_2=\pi/10$. }
\end{figure}

The asymptotic variances of $\phi_k$ ($k=1,2$ for two ligands respectively) can be obtained by the following inequality \cite{distri}:
\begin{equation}
\textbf{C}_k-\textbf{R}^T_k\textbf{F}^{-1}\textbf{R}_k\geq{0}, \label{eq:14}
\end{equation}
where $\textbf{R}_k$ is the transformation vector of $\phi_k$, $\textbf{F}^{-1}$ is the inverse of information matrix. $\textbf{C}_k$ is the minimum fluctuation of $\phi_k$. The definition of $\textbf{R}_k^T$ is:
\begin{eqnarray}
\textbf{R}_k^T = \left[
\begin{array}{cccc}
\frac{\partial{\phi_k}}{\partial{\overline{x}_1}} &\frac{\partial{\phi_k}}{\partial{\overline{x}_2}} & \frac{\partial{\phi_k}}{\partial{\overline{x}_3}} & \frac{\partial{\phi_k}}{\partial{\overline{x}_4}} \\
\end{array}\right].\label{eq:15}
\end{eqnarray}
By expanding the $C_n$ and $D_n$ to second order of $p$, we obtain $\phi_1$ and $\phi_2$ as:
\begin{subequations} \label{eq:16}
\begin{equation}
\phi_1=\arctan[\frac{(2+\frac{1}{b_1})\overline{x}_2-(1-\frac{1}{b_1})\overline{x}_4}{(2+\frac{1}{b_1})\overline{x}_1-(1-\frac{1}{b_1})\overline{x}_3}],
 \label{subeq:1601}
\end{equation}
\begin{equation}
\phi_2=\arctan[\frac{(1-\frac{1}{b_2})\overline{x}_2-(2+\frac{1}{b_2})\overline{x}_4}{(1-\frac{1}{b_2})\overline{x}_1-(2+\frac{1}{b_2})\overline{x}_3}]. \label{subeq:1602}
\end{equation}
\end{subequations}
The minimum of $\textbf{C}_k$ is obtained by inequality [Eq. (\ref{eq:14})]. By taking $p_2=p_1=p$, FIG. \ref{fig:seven} and \ref{fig:eight} are the results of fixed $b_1$ with different $b_2$ by numerical differentiating Eq. (\ref{eq:16}) with statistical parameters $x_i$.

One would expect that small asymptotic variance presents better sensing ability of the ligand. Due to the lack of experimental evidence for $\sigma_{\phi}^2$, we set a criteria $\sigma_{\phi}^2\leq1$ to be the condition for sensing ability. In FIG. \ref{fig:seven}, with $b_1=0.01$, $b_2=0.4$ and $p=0.02$, $\sigma_{\phi_1}^2\simeq300$ and $\sigma_{\phi_2}^2\simeq1000$, it seems that the cell can not have a good sensing to any ligand. As $p$ increases, for example $p=0.4$, $\sigma_{\phi_1}^2=0.81864$ and $\sigma_{\phi_2}^2=2.50598$, which means the cell has small asymptotic variances for better resolution of the ligands. One may infer that the cell might recognize a more distinct direction and move toward Ligand 1. Meanwhile, the difference of the asymptotic variances are not so eminent when these $b_j$ ($j=1,2$) are almost equal, this can be seen from FIG. \ref{fig:seven} that for the case $b_2=0.02$ and $p=0.61$, $\sigma_{\phi_1}^2=0.98659$ and $\sigma_{\phi_2}^2=1.01726$, and the cell could not determine a prefer direction for motion, even though $\sigma_{\phi}^2$ satisfy our criteria but the cell might be undecidable due to $\sigma_{\phi_1}^2\simeq\sigma_{\phi_2}^2$.

It is interesting to see whether the above observation depends on the magnitude of $b_j$. The results from fixed $b_1=100$ with different $b_2$ are plotted in FIG. \ref{fig:eight}. For the case of $b_2=400$, we have $\sigma_{\phi_1}^2=14.99085$ and $\sigma_{\phi_2}^2=187.88$ at $p=0.04$, and $\sigma_{\phi_1}^2=0.95067$ and $\sigma_{\phi_2}^2=11.91643$ for $p=0.16$.
It seems that better sensing resolution exhibits with larger gradient. On the other hand, for the case $b_1=100$, $b_2=120$ and $p=0.2$, $\sigma_{\phi_1}^2=0.812736$ and $\sigma_{\phi_2}^2=1.1280$, confusing information also exists in this situation. Therefore, FIG. \ref{fig:eight} indicates the occurrence of good resolution with large difference of $b_j$ is a general feature in this theoretical model. However, further experimental observation will be needed to clarify this assertion.

\begin{figure}
\includegraphics[width=0.4\textwidth]{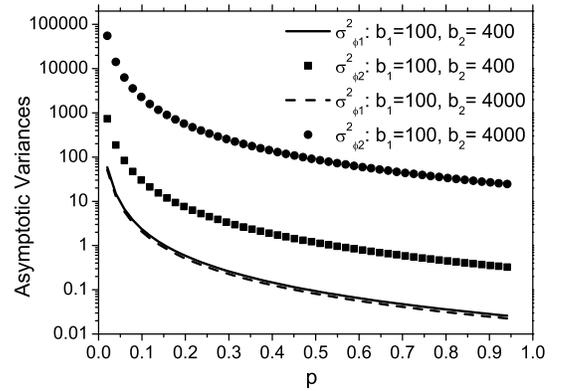}\\% Here is how to import EPS art
\caption{\label{fig:nine} The results of $\sigma_{\phi_1}^2$ and $\sigma_{\phi_2}^2$ versus $p$ for $b_1=100$ and $b_2=400, 4000$ with $\phi_1=0$ and $\phi_2=\pi/10$. }
\end{figure}

Since large $b_j$ corresponds to unbinding, it is important to see the effect of unbinding on FIG. \ref{fig:nine}. For large $b_2$, such as $b_2=4000$, the results show the cell has no information for Ligand 2 through binding, therefore one would expect a large $\sigma_{\phi_2}^2$ and our results in FIG. \ref{fig:nine} also shows such tendency. Furthermore, when $b_2$ decrease to $400$, the cell can have more binding receptors ,therefore $\sigma_{\phi_2}^2$ decreases.

Above discussion indicated better sensing resolution exists when both conditions $\sigma_{\phi}^2\leq1$ and large difference of $\sigma_{\phi_1}^2$ and $\sigma_{\phi_2}^2$ are satisfied. In addition, the asymptotic variance decreases when the receptor tends to bind with ligand and cell could receive more information from ligands.

We have provided the extension to deal with two ligands system by using Michaelis-Menten model. Under the equilibrium situation, minimum fluctuations of sensing ability can be obtained. It is noted that such system can be analyzed by including the ligand concentrations inside the partition function.

\section{\label{sec:level3}CONCLUSION}
In this work we have modified the mechanism of sensing ability by including the dynamics of the ligand. In our approach we are able
to avoid having energies correlate with the concentration of ligand. This was accomplished by setting up the system with different energy levels and chemical potentials, and use grand canonical partition function to address the sensing ability. It is interesting that this model can still exhibit remarkable sensing ability. Moreover, in our approach the energy levels are free parameters which can be used for any cellular complex, therefore our model has predicting power for other physical quantities such that further experimental results can be used to justify the correctness of this kind of model.

We have also studied in more complicate environment with multiple ligands under the competitively binding situation. The results indicate that cells can distinguish different ligands under the small gradient with different chemical dissociation constants. This extension gives the possible way to predict the reactions of the cell under the practical biological environment.

\begin{acknowledgments}
This work was supported by National Science Council of Taiwan, NSC-97-2112-M-006-003-MY2. S. H. Liou thanks the support from
National Center for Theoretical Science, Taiwan. We would also like to thank our referees for the suggestion which formed the results of section IV.
\end{acknowledgments}

\end{document}